\documentclass[aps,reprint,prb,superscriptaddress,amsmath,amssymb,floatfix]{revtex4-2}

\usepackage[utf8]{inputenc}
\usepackage{graphicx}
\usepackage[colorlinks,allcolors=blue]{hyperref}
\usepackage{braket}

\begin{document}

\title{Multifractality in the interacting disordered Tavis-Cummings model}

\author{F.~Mattiotti}
\affiliation{University of Strasbourg and CNRS, CESQ and ISIS (UMR 7006), aQCess, 67000 Strasbourg, France}

\author{J.~Dubail}
\affiliation{University of Strasbourg and CNRS, CESQ and ISIS (UMR 7006), aQCess, 67000 Strasbourg, France}
\affiliation{Universit\'{e} de Lorraine and CNRS, LPCT (UMR 7019), 54000 Nancy, France}

\author{D.~Hagenm\"uller}
\affiliation{University of Strasbourg and CNRS, CESQ and ISIS (UMR 7006), aQCess, 67000 Strasbourg, France}

\author{J.~Schachenmayer}
\affiliation{University of Strasbourg and CNRS, CESQ and ISIS (UMR 7006), aQCess, 67000 Strasbourg, France}

\author{J.-P.~Brantut}
\affiliation{Ecole Polytechnique F\'ed\'erale de Lausanne, Institute of Physics, CH-1015 Lausanne, Switzerland}
\affiliation{Center for Quantum Science and Engineering, Ecole Polytechnique F\'ed\'erale de Lausanne, CH-1015 Lausanne, Switzerland}

\author{G.~Pupillo}
\affiliation{University of Strasbourg and CNRS, CESQ and ISIS (UMR 7006), aQCess, 67000 Strasbourg, France}
\affiliation{Institut Universitaire de France (IUF), 75000 Paris, France}

\begin{abstract}
We analyze the spectral and transport properties of the interacting disordered Tavis-Cummings model at half excitation filling. We demonstrate that a Poissonian level statistics coexists with eigenfunctions that are multifractal (extended, but non-ergodic) in the Hilbert space, for all strengths of light-matter interactions. This is associated with a lack of thermalization for a local perturbation. We find that the bipartite entanglement entropy grows logarithmically with time, similarly to many-body localized systems, while the spin imbalance tends to zero for strong coupling, in analogy to ergodic phases. We show that these effects are due to the combination of finite interactions and integrability of the model. When a small integrability-breaking perturbation (nearest-neighbor hopping) is introduced, typical eigenfunctions become ergodic, seemingly turning the system into a near-perfect conductor, contrary to the single-excitation noninteracting case. We propose a realization of this model with cold atoms.
\end{abstract}

\maketitle

\section{Introduction}

While fractal objects are usually characterized by a single fractal dimension, multifractality extends this concept to scale-invariant probability distributions characterized by a continuous spectrum of fractal dimensions~\cite{halsey1986fractal}. Multifractality plays an important role in a wide variety of situations including earth topography~\cite{schertzerNPG2006}, turbulent flows~\cite{meneveau_sreenivasan_1991}, heartbeat dynamics~\cite{Ivanov1999}, and financial markets~\cite{Mandel1999}. Another example is the critical point of Anderson metal-insulator transitions in disordered quantum systems~\cite{Anderson1958}, at which adjacent energy levels typically follow a semi-Poisson distribution~\cite{shorePRB1993,schmitPRE1999,schmitPRL2004}, and where the multifractal spectrum identifies the universality class of the transition~\cite{Mirlin2008}. Multifractality reflects the extended, yet nonergodic, nature of the eigenstates, which is usually associated with finite but limited conductivity. This is in contrast to both good conductors (fully ergodic eigenstates, Wigner-Dyson level statistics) and insulating phases (fully nonergodic eigenstates, Poisson level statistics)~\cite{Mirlin2008}. Beyond single-particle wavefunctions, multifractal behavior has been reported in strongly interacting disordered quantum systems in connection with many-body localization (MBL)~\cite{torres2015dynamics,laflorenciePRL2019,mirlinPRB2018,BarLev2020,de2021rare}, and in certain random matrix models~\cite{schmitPRE1999,monthus2017multifractality,ioffePRR2020,tarziaPRB2021,backer2019multifractal}, including the Rosenzweig-Porter model~\cite{aminiNJP2015,biroliE2016,ossipovE2016,sieberPRE2018}, Bethe-ansatz integrable models~\cite{shastryNJP2016,khaymovichSP2022}, models with correlated hopping~\cite{borgonoviPRB2016a,santosPRL2016,santosPRL2018a,kravtsovPRB2019a,kremsPRB2018,khaymovichSP2021} or diagonal disorder~\cite{gopalakrishnanPRB2017,santosPRL2019,sharmaPRB2021,flachPRB2023,flachPRB2023a}, and one-dimensional (1D) quantum circuits~\cite{turkeshiPRL2022}. The precise connection between interactions, multifractality, level statistics, and transport properties, as well as experimental realizations of multifractal behavior in any interacting quantum many-body systems, have so far remained largely elusive.

\begin{figure}[t]
    \centering
    \includegraphics[width=\columnwidth]{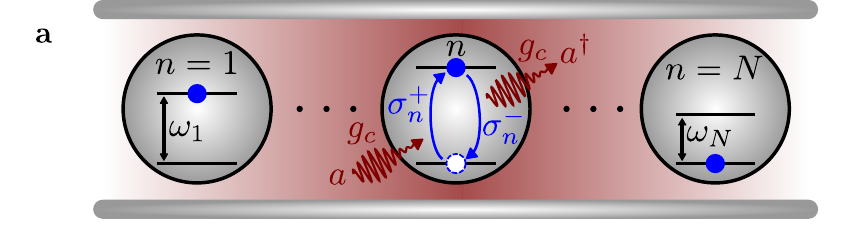}
    \includegraphics[width=\columnwidth]{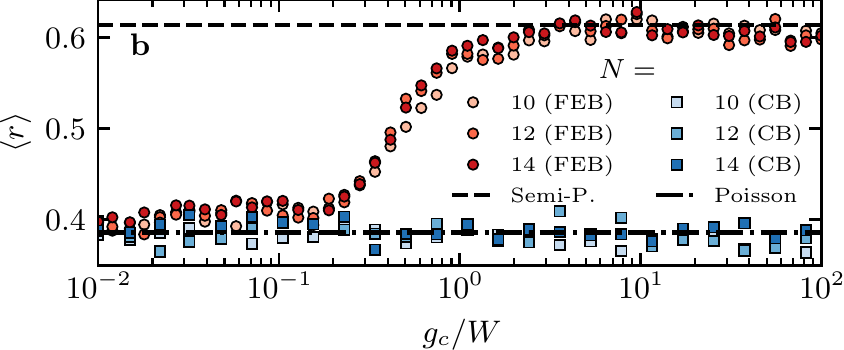}
    \caption{\textit{Level statistics in the disordered TC model}. {\bf a} $N$ two-level systems with transition energies $\omega_n \in [-W/2,W/2]$ ($W$ is the disorder strength) are coupled to a cavity mode with collective coupling strength $g_c$. The total number of excitations, which can be either in the spins (blue, $\sigma^{\pm}_{n}$) or in the cavity (red, $a^{(\dagger)}$) is denoted as $M$. {\bf b} The mean level spacing ratio [Eq.~\eqref{eq:r}] for the first-excited band (FEB) and the central band (CB) is shown as a function of $g_c$ at half filling~\cite{Note1}.}
    \label{fig:DOS}
\end{figure}

In this work we investigate the disordered Tavis-Cummings (TC) model of cavity quantum electrodynamics~\cite{tavis1968exact} that describes $N$ two-level systems---or pseudospins---coupled to a single cavity mode (Fig.~\ref{fig:DOS}{\bf a}). We consider diagonal disorder in the spin energies typically associated to inhomogeneous broadening~\cite{temnov2005superradiance,szymanska2006nonequilibrium,diniz2011strongly,belyaninPRA2023}. The total number of excitations $M$ is conserved in the TC model; here we focus on half filling $(M=N/2)$ and numerically demonstrate that a multifractal spectrum exists in the computational basis for all strengths of the light-matter coupling. In contrast to the single-excitation case, featuring multifractality, semi-Poisson level statistics, and diffusivelike transport~\cite{schachenmayerPRB2020,hagenmullerPRA2022}, here we simultaneously observe Poisson statistics and multifractality in the many-body Hilbert space (a similar phenomenon has been pointed out recently in certain power-law banded random matrix models~\cite{khaymovichQ2022}). We demonstrate that interactions tend to favor delocalization in the Hilbert space, and analyze the thermalization of a single-spin excitation with the microcanonical Edwards-Anderson (EA) order parameter introduced in Ref.~\cite{scardicchioPRB2011}. We find a nonvanishing EA parameter, signaling the absence of thermalization, for all light-matter coupling strengths. We also find that the bipartite entanglement entropy grows logarithmically with time, and the spin imbalance vanishes for strong coupling. Interestingly, our findings show that the disordered TC model possesses various known features of MBL phases, including Poisson statistics~\cite{HuseMBL2010,ScardicchioMBL2017}, lack of thermalization~\cite{PolyakovMBL2005,AltshulerMBL2006}, multifractality~\cite{AletMBL2015,laflorenciePRL2019,BarLev2020}, and logarithmical growth of entanglement entropy in time~\cite{aletPRB2016,ScardicchioMBL2017}. However, unlike MBL, the spin imbalance goes to zero~\cite{aletPRB2016,blochS2015} and, importantly, these features are found to be unstable towards a small perturbation, and they originate instead from the integrability of the disordered TC model. We show this by adding a weak nearest-neighbor hopping term to the Hamiltonian, which turns the multifractal phase into a perfect (ergodic) conductor for typical energies in the middle of the spectrum, while the low-energy states remain multifractal. This is in contrast to the single-particle case, which is robust with respect to this perturbation~\cite{schachenmayerPRB2020}. We propose a protocol to experimentally analyze this physics with ultracold atoms.

\section{Model}

The Hamiltonian ($\hbar=1$) reads
\begin{equation}
\label{eq:h}
    H_{\rm TC} = \omega a^\dag a + \sum_{n=1}^N \omega_n \sigma_n^+\sigma_n^- + \frac{g_c}{\sqrt{N}} \sum_{n=1}^N \left( a^\dagger \sigma_n^- + \sigma_n^+ a \right)~,
\end{equation}
where $\omega$ is the cavity frequency, $\omega_n$ are random energies, $g_c$ is the collective coupling strength of the emitters to the cavity mode, $\sigma_n^\pm$ are the Pauli ladder operators, and $a^\dagger/a$ is the bosonic creation/annihilation operator of a cavity photon. Since the total number of excitations
\begin{equation}
    M = a^{\dagger} a + \sum_{n} \sigma^+_n\sigma^-_n
\end{equation}
is conserved, the spectrum is fully determined by the detunings \mbox{$\omega_n-\omega$} and  $g_c$, and without loss of generality we can set \mbox{$\omega=0$.} In this Letter we consider random, uniformly distributed detunings $\omega_n \in [-W/2,+W/2]$. The coupling between spin-1/2 operators and the bosonic mode makes the Hamiltonian Eq.~\eqref{eq:h} nonquadratic and interacting. The (non-normalized) eigenstates of $H$ are of the form~\cite{gaudin1976diagonalisation,tschirhart2014algebraic,tschirhart2017,claeys2018richardson,tschirhart2018steady,claeysSP2023} \begin{equation}
    \ket{\lambda_1,\dots,\lambda_M} = \Sigma^\dag(\lambda_1)\cdots \Sigma^\dag(\lambda_M) \ket{0}~,
\end{equation}
where the raising operators
\begin{equation}
    \Sigma^\dag(\lambda) = a^\dag + \frac{1}{\sqrt{N}} \sum_{n=1}^N \frac{g_c}{\lambda - \omega_n} \sigma^+_n
\end{equation}
act on the vacuum state $\ket{0}$ (no exciton and no photon). The energy of the eigenstate is
\begin{equation}
    E=\sum_{j=1}^M \lambda_j~.
\end{equation}
The coefficients $\lambda_1,\dots,\lambda_M$, called ``Bethe roots,'' satisfy a system of $M$ coupled nonlinear equations
\begin{equation}
    \label{eq:bethe}
    -\frac{N\lambda_i}{2g_c^2} + \frac{1}{2} \sum_{n=1}^N \frac{1}{\lambda_i-\omega_n} = \sum_{\substack{j=1 \\ j\neq i}}^M \frac{1}{\lambda_i-\lambda_j},
\end{equation}
called ``Bethe equations,'' which are a consequence of the underlying integrability of the model~\cite{gaudin1976diagonalisation,tschirhart2014algebraic,tschirhart2017,claeys2018richardson,tschirhart2018steady,claeysSP2023}.

\section{Spectral properties}

\subsection{Energy bands}

\begin{figure}[b]
    \centering
    \includegraphics[width=\columnwidth]{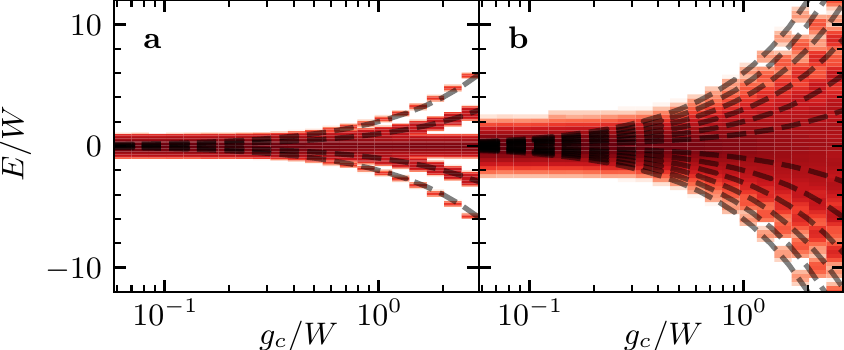}
    \caption{\textit{Energy bands in the disordered TC model}. Density of states of the TC model for $N=12$ as a function of the energy $E$ and the coupling $g_{c}$ for $M=2$ ({\bf a}) and $M=N/2=6$ ({\bf b}). Dashed lines: Low-excitation density approximation ($M \ll N$) $E=j g_c$, with $j=-M,\dots,M$.}
    \label{fig:DOS2}
\end{figure}

For a low excitation density $M \ll N$, the right-hand side in Eq.~(\ref{eq:bethe}) is negligible, and the Bethe roots approximately coincide with the eigenvalues of an arrowhead matrix~\cite{hagenmullerPRA2022}. For large coupling $g_c \gg W$, these correspond to
\begin{subequations}
    \begin{enumerate}
        \item two polaritons
        \begin{equation}
            \lambda_i \simeq \pm g_c~,
        \end{equation}
        and
        \item $N-1$ dark states
        \begin{equation}
            \lambda_i \simeq \Omega_n :=  \frac{\omega_{n+1}+\omega_n}{2}~,
        \end{equation}
    \end{enumerate}
\end{subequations}
where we assume without loss of generality that $\omega_1< \omega_2 < \dots < \omega_N$. For instance, for $M=2$ excitations the approximate basis is formed by: (i) combinations of two polaritons, with energy $E\sim \pm 2g_c$ and $E\sim -g_c+g_c=0$; (ii) combinations of a polariton and a dark state $E\sim\pm g_c+\Omega_n$; (iii) combinations of two different dark states, $E\sim\Omega_{n_1}+\Omega_{n_2}$. For $M$ excitations and strong coupling, there are $2M+1$ energy bands centered at $E\sim j g_c$, with $j=-M,\dots,M$. In Fig.~\ref{fig:DOS2}, we compare the low-density approximation (dashed lines) to the numerical density of states for $N=12$. While the low-density approximation works well for $M=2$ (Fig.~\ref{fig:DOS2}{\bf a}), it fails at half filling ($M=6$, Fig.~\ref{fig:DOS2}{\bf b}), which is expected since the interaction effects become strong in the latter case. Nevertheless, the low-density approximation provides an intuitive characterization of the many-body eigenstates for strong coupling, based on their energies. From now on, \emph{we only focus our analysis on half filling ($M=N/2$), and consider three sets of states as representatives of the full Hilbert space}:
\begin{subequations}
    \begin{enumerate}
        \item the ground state (GS)
        \begin{equation}
            E\sim -Mg_c~,
        \end{equation}
        \item the first excited band (FEB)
        \begin{equation}
            E\sim -(M-1)g_c+\Omega_n
        \end{equation}
        for $n=1,\dots,N-1$, and
        \item the central band (CB)
        \begin{equation}
            E\sim\sum_{j=1}^M\Omega_{n_j}
        \end{equation}
        for sets of $M$ mutually distinct indices $n_j$.
    \end{enumerate}
\end{subequations}
In our numerical analysis, for the CB we select a small energy window made of the $N$ states in the middle of the spectrum. Note that the results depend on the energy bands when the latter are well separated. For large $M$ the width of the CB increases as $\sim\sqrt{M}$ and so it overlaps with the neighboring bands and the low-density description fails. Nevertheless, even in that limit, the specific energy bands that we are considering are only weakly modified, since the GS and the FEB remain gapped, and the vast majority of the states at $E\sim 0$ are of the form of the CB defined above.

\subsection{Level statistics}

We first analyze how the statistics of the energy levels depends on the coupling strength $g_c$---the statistics are usually related to the localization properties of many-body interacting systems~\cite{santosPRL2020,laflorencieCRP2018}---by computing the mean level spacing ratio~\cite{husePRB2007,laflorencieCRP2018,santosPRL2020}
\begin{equation}
\label{eq:r}
    \braket{r}=\left\langle \frac{\min (E_{\alpha+1}-E_{\alpha},E_{\alpha}-E_{{\alpha}-1})}{\max (E_{\alpha+1}-E_{\alpha},E_{\alpha}-E_{{\alpha}-1})}\right\rangle~,
\end{equation}
with $E_{\alpha}$ the sorted eigenenergies. The mean level spacing ratio is shown in Fig.~\ref{fig:DOS}{\bf b} as a function of $g_c$ (FEB: circles, CB: squares). As we discuss below, the CB is representative of many other energy bands, while the FEB has peculiar features. We find that the level statistics is Poissonian for small coupling for both bands, while for large coupling $g_c \gg W$, the FEB reaches semi-Poissonian statistics and the CB remains Poissonian.

These differences can be understood using the low-density description of the many-body eigenstates: the states of the FEB have an energy $E\sim -(M-1)g_c+\Omega_n$, and thus keep the semi-Poisson statistics of the dark states in the single-excitation regime~\cite{schachenmayerPRB2020,hagenmullerPRA2022}. On the other hand, the states of the CB have an energy $E\sim \Omega_1+\Omega_2+\dots+\Omega_M$, and since they result from the sum of independent random single-excitation energies, they do not exhibit level repulsion (Poisson level statistics). We have checked that other energy bands formed by more than one dark state also follow Poisson statistics. The Poisson level statistics is consistent with the integrability of the model: the extensive number of conserved charges prevents level repulsion in the center of the spectrum.

\subsection{Multifractality}

\begin{figure}[b]
    \centering
    \includegraphics{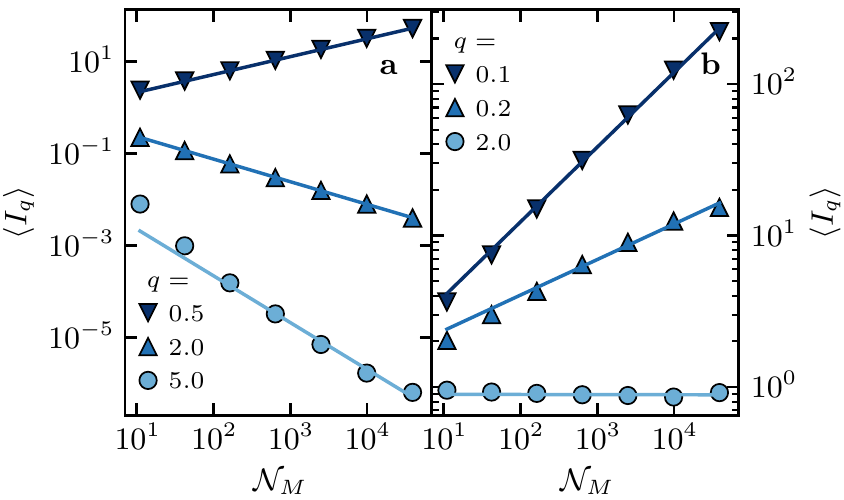}
    \caption{\textit{Multifractality of the wavefunctions for strong ({\bf a}, $g_c/W=10$) and weak coupling ({\bf b}, $g_c/W=0.01$)}. Finite size scaling of $\braket{I_q}$ [Eq.~\eqref{eq:ipr}] ($N$ is varied between 4 and 16) in the CB for different $q$. Lines: power-law fits $\propto {\cal N}_M^{-\tau_q}$.}
    \label{fig:IPRq1}
\end{figure}

We now proceed to analyze the delocalization of the many-body eigenfunctions in the Hilbert space by studying their multifractality. We define the generalized inverse participation ratio (IPR) in the many-body case (averaged in an energy window) as
\begin{equation}
\label{eq:ipr}
    \langle I_q \rangle = \left\langle \sum_{p=1}^{\mathcal{N}_M} |\langle p | E_\alpha \rangle|^{2q} \right\rangle,
\end{equation}
with
\begin{equation}
    \ket{p}=(a^\dag)^{n_p}\prod_{j=1}^{n_s}\sigma_j^+\ket{0}
\end{equation}
the many-excitation product states with $n_p$ photons and $n_s$ excitons ($n_s+n_p=M$) that form an eigenbasis for $g_c=0$, and ${\cal N}_M=\sum_{i=0}^M \binom{N}{i}$ the size of the Hilbert space subset with $M$ excitations. The average $I_q$ is typically a power-law function of the Hilbert space size $\langle I_q \rangle \propto {\cal N}_M^{-\tau_q}$, where the exponent $\tau_q$ characterizes the extent of the wavefunction~\cite{laflorenciePRL2019,de2021rare,hagenmullerPRA2022,Mirlin2008}. While $|\langle p | E_\alpha \rangle|^2 \sim O({\cal N}_M^{-1})$ and thus $\tau_q = q-1$ for fully extended (ergodic) eigenfunctions, localized states feature one or few components $|\langle p | E_\alpha \rangle|^2 \sim O(1)$, implying $\tau_q=0$. Any non-linear dependence on $q$ identifies the system as multifractal~\cite{Mirlin2008}.

\begin{figure}[t]
    \centering
    \includegraphics[width=\columnwidth]{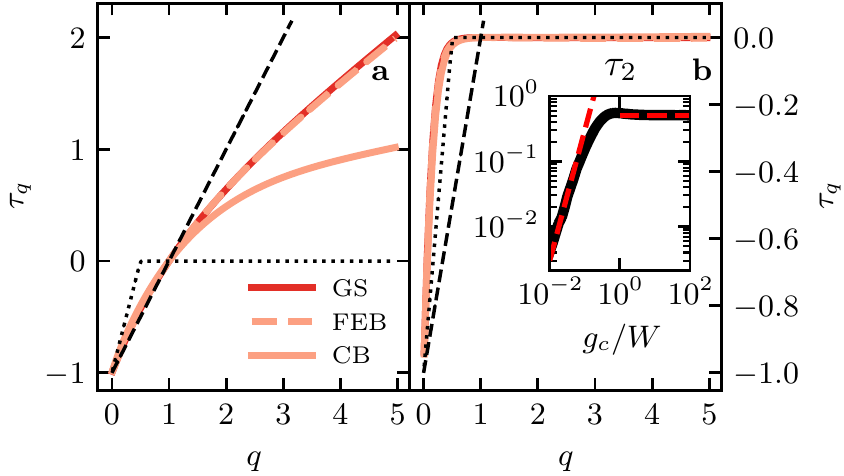}
    \caption{\textit{Multifractality of the wavefunctions for strong ({\bf a}, $g_c/W=10$) and weak coupling ({\bf b}, $g_c/W=0.01$)}. Fitted exponent $\tau_q$ (obtained from $\braket{I_q}\propto {\cal N}_M^{-\tau_q}$ (see Fig.~\ref{fig:IPRq1}) as a function of $q$ for the GS, FEB, and CB. Data for GS and FEB overlap in {\bf a}, while all curves overlap in {\bf b}. Black dashed line: fully ergodic case $\tau_q=q-1$; Dotted line: single-excitation case $\tau_q=\min(2q-1,0)$~\cite{hagenmullerPRA2022}. Inset: $\tau_2$ (averaged over the full spectrum) versus $g_c$. Dashed lines: asymptotic behaviors $\tau_2 \sim (g_c/W)^2$ and $\tau_2\sim 0.5$.}
    \label{fig:IPRq}
\end{figure}

In Fig.~\ref{fig:IPRq1} we show that the generalized IPR $\braket{I_q}$ [Eq.~\eqref{eq:ipr}] has a power-law dependence on the Hilbert space size ${\cal N}_M$ for any $q$ and both for strong and weak coupling to the cavity ({\bf a} and {\bf b}, respectively). This allows us to determine the multifractal exponent $\tau_q$ from power-law fits $\braket{I_q}\propto {\cal N}_M^{-\tau_q}$. The exponent $\tau_q$ obtained by fitting the finite-size scaling of $\langle I_q \rangle$ is shown in Fig.~\ref{fig:IPRq}. The large deviations from the ergodic case (dashed line) indicate a clear multifractal behavior for both strong ({\bf a}) and weak ({\bf b}) coupling, and for all the energy bands. For strong coupling, $\tau_q$ grows monotonically with $q$, similarly to what is observed in power-law banded random matrix models~\cite{Mirlin2008}. As the coupling decreases, $\tau_q$ approaches the single-excitation (noninteracting) case~\cite{hagenmullerPRA2022} (dotted line) for all the energy bands, with $\tau_q\approx 0$ for $q>1/2$. Such a behavior usually characterizes a transition to a ``frozen phase,'' that combines properties of localized and critical states~\cite{Mirlin2008}. However, here multifractality varies smoothly with the coupling strength, with no clear sign of a phase transition. This is shown in the inset of Fig.~\ref{fig:IPRq}{\bf b}, where $\tau_2$ (related to the standard IPR) averaged over the full spectrum is shown as a function of $g_c/W$. For weak coupling, we observe a smooth dependence $\tau_2\sim (g_c/W)^2$ (diagonal dashed line), which can be explained via resonance counting~\cite{mirlinPRB2018,mirlinAoP2021,burinAdP2017}: indeed, there is a probability $\sim (g_c/W)^2/N$ that two spins are simultaneously resonant with the cavity, which leads to a hybridization, providing a factor $\sim 1/2$ to the IPR. Since we have $O(N)$ excitations for each many-body state, we have $\sim N^2$ terms, so that 
\begin{equation}
    \braket{I_2}\sim 2^{-N(g_c/W)^2} \qquad \text{for } g_c \ll W.
\end{equation}
The Hilbert space size scales as ${\cal N}_M \sim 2^N$, implying $\tau_2 \sim (g_c/W)^2$. The first-order term, i.e., the contribution from the hybridization of single spins with the cavity, provides a contribution $(g_c/W)/\sqrt{N}$, which vanishes in the $N \to \infty$ limit. As the coupling increases, $\tau_2$ grows and saturates to a finite value $\tau_2\approx 0.5 < 1$ (horizontal dashed line in the inset of Fig.~\ref{fig:IPRq}{\bf b}), indicating that the eigenstates are multifractal for any coupling strength. As we show in the following, this is related to the integrability of the model and it is in contrast to models showing MBL, where $\tau_2$ is finite and nonunity only for strong disorder~\cite{laflorenciePRL2019}. Interestingly, here the eigenfunctions for strong coupling are more delocalized than in the noninteracting case, where multifractality has been associated to diffusivelike transport~\cite{schachenmayerPRB2020,hagenmullerPRA2022}.

\section{Observables}

\subsection{Transport}

\begin{figure}
    \centering
    \includegraphics[width=\columnwidth]{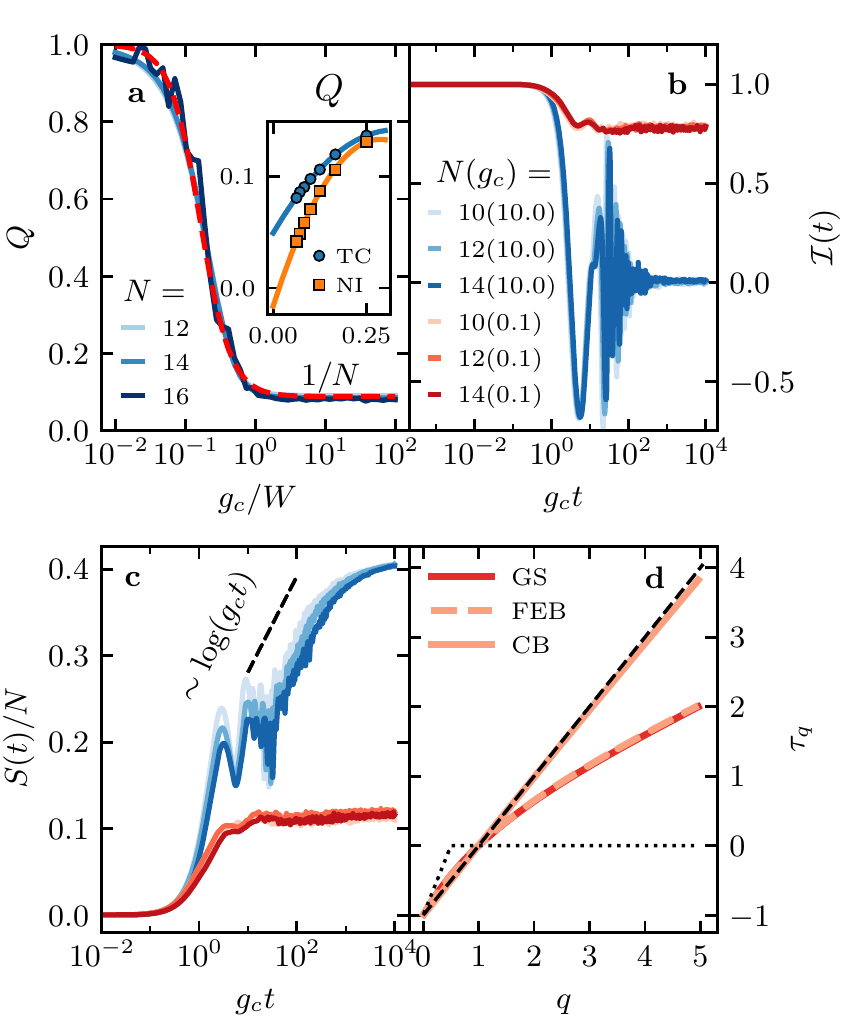}
    \caption{\textit{Absence of thermalization, dynamics, and sensitivity to integrability breaking}. {\bf a} Survival probability $Q$ [Eq.~\eqref{eq:EApar}] as a function of $g_c$. Dashed line: best fit $Q=[1+A(g_c/W)^2]/[1+B(g_c/W)^2]$, with $A=3.3$ and $B=37.4$. Inset: Finite-size scaling $Q(N)=Q + c_1/N+c_2/N^2$ (continuous lines), both in the integrable case [TC: Eq.~\eqref{eq:h}, $g_c/W=10$] and in the non-integrable case [NI: Eq.~\eqref{eq:hni}, $g_c/W=10$, $J/W=0.1$]. {\bf b,c}~Spin imbalance ({\bf b}) [Eq.~\eqref{eq:imb}] and bipartite entanglement entropy ({\bf c}) [Eq.~\eqref{eq:ent}] as a function of time $t$ (scaled by $g_c$, with $W=1$) starting from the Néel state $\prod_{j=1}^{N/2} \sigma_{2j}^+ \ket{0}$. {\bf d}~Fitted exponent $\tau_q$ as a function of $q$ in the nonintegrable model ($g_c/W=10$, $J/W=0.1$). Data for GS and FEB overlap. Black dashed and dotted lines are the same as in Fig.~\ref{fig:IPRq}.}
    \label{fig:EApar}
\end{figure}

To characterize the transport properties of the model, we now analyze the long-time survival probability, or microcanonical EA parameter, introduced in Ref.~\cite{scardicchioPRB2011}:
\begin{equation}
\label{eq:EApar}
    Q = \frac{1}{{\cal N}_M} \sum_{\alpha=1}^{{\cal N}_M} \frac{1}{N}\sum_{n=1}^N \vert \braket{E_{\alpha} \vert \sigma_n^z \vert E_{\alpha}}\vert^2~.
\end{equation}
This parameter has a simple physical interpretation~\cite{scardicchioPRB2011}. Let the system be initialized in an infinite-temperature state with a small magnetization of spin $n$, i.e.,
\begin{equation}
    \braket{\sigma_n^z}_0={\rm Tr}(\rho_0 \sigma_n^z)=\epsilon~,
\end{equation}
with the density matrix $\rho_0=(\mathbb{I}+\epsilon\sigma_n^z)/{\cal N}_M$. Then, the long-time local magnetization is \begin{align}
    \braket{\sigma_n^z}_\infty &=\lim_{t\to\infty}{\rm Tr}(e^{-iHt}\rho_0e^{iHt} \sigma_n^z) \nonumber \\
    &=\frac{\epsilon}{{\cal N}_M} \sum_{\alpha=1}^{{\cal N}_M} \vert \braket{E_{\alpha} \vert \sigma_n^z \vert E_{\alpha}}\vert^2
\end{align}
and Eq.~\eqref{eq:EApar} thus represents the ratio of the long-time local magnetization over the initial one $\braket{\sigma_n^z}_\infty/\braket{\sigma_n^z}_0$, averaged over all the spins. In the Richardson model---a related many-body spin model without cavity mode and with all-to-all couplings---it was shown that $Q$ remains finite for any coupling strength for $N \rightarrow \infty$, and vanishes only at infinite coupling~\cite{scardicchioPRB2011}. This absence of thermalization at finite coupling was related to the nonergodicity of the eigenfunctions. Here we perform the same analysis in the TC model and observe that $Q$ saturates to a finite value upon increasing $N$, for any finite $g_c$ (see Fig.~\ref{fig:EApar}{\bf a}). While this value is close to unity for small coupling, indicating a nearly perfect insulator, it decreases with $g_c$. In contrast to the Richardson model, however, we observe that $Q$ approaches a positive constant asymptotically as $g_{c}/W \to \infty$, thus indicating an absence of thermalization even at infinite coupling.

\subsection{Dynamics}

While the steady-state properties of the TC model are similar to the MBL phase~\cite{laflorencieCRP2018,laflorenciePRL2019,ScardicchioMBL2017}, the dynamics reveals a more peculiar behavior. To show this, we initialize the system in the Néel state (with random $\omega_n$) and we first look at the time evolution of the spin imbalance~\cite{blochS2015,aletPRB2016}
\begin{equation}
    \label{eq:imb}
    {\cal I}(t) = \frac{1}{N} \sum_{n=1}^N (-1)^n \braket{\psi(t) \vert \sigma_n^z \vert \psi(t)}~,
\end{equation}
shown in Fig.~\ref{fig:EApar}{\bf b}, which reaches zero at long time for large $g_c$, indicating ergodicity. In contrast, for weak $g_c$ it saturates to a finite value, as in MBL~\cite{aletPRB2016}. However, we observe that the von Neumann bipartite entanglement entropy
\begin{equation}
    \label{eq:ent}
    S(t) = -{\rm Tr}\left[ \rho_{N/2}(t) \log_2 \rho_{N/2}(t) \right]~,
\end{equation}
shown in Fig.~\ref{fig:EApar}{\bf c} (where $\rho_{N/2}$ is the trace of $\ket{\psi(t)}\bra{\psi(t)}$ over a set of $N/2$ spins randomly chosen), always grows logarithmically at long times, similar to the MBL phase of XXZ chains~\cite{aletPRB2016,ScardicchioMBL2017}. The prefactor depends on $g_c/W$ and it appears negligible for $g_c/W=0.1$, compared to $g_c/W=10$. Moreover, by dividing $S(t)/N$, there seems to be a scaling $S(t)\sim N$, which is in contrast to the area law observed in MBL.

\section{Integrability-breaking perturbations}

\begin{figure}[b]
    \centering
    \includegraphics{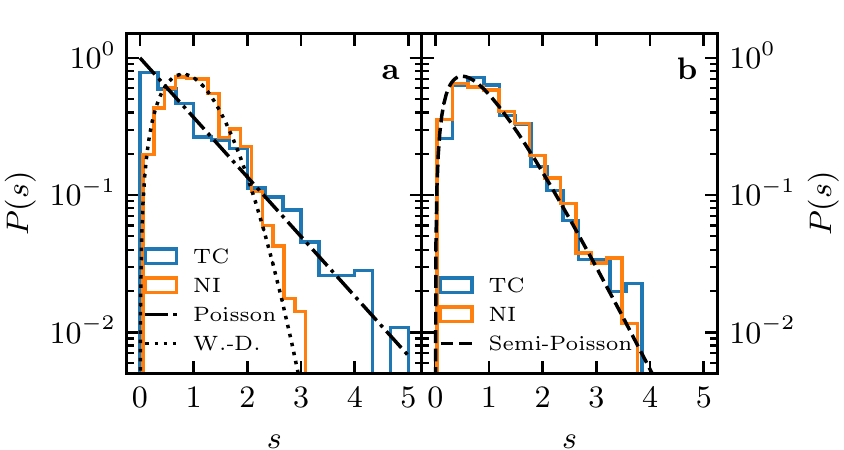}
    \caption{\textit{Sensitivity to integrability breaking}. Probability density $P(s)$ of the normalized level spacing $s$ in the CB ({\bf a}) and the FEB ({\bf b}) for $N=14$, both in the integrable (TC) and non-integrable (NI) cases. ``Poisson'': $P(s)=e^{-s}$, ``Semi-Poisson'': $P(s)=4se^{-2s}$, ``Wigner-Dyson'' (W.-D.): $P(s)=\frac{\pi}{2}se^{-\pi s^2/4}$. Parameters: $g_c/W=10$ and $J/W=0.1$.}
    \label{fig:Ps}
\end{figure}

An important question is now if the TC model features survive integrability-breaking perturbations. To answer this, we add a small nearest-neighbor hopping term
\begin{equation}
\label{eq:hni}
    H_{\rm NI} = H_{\rm TC} + J \sum_{n=1}^{N-1} \left( \sigma_n^+\sigma_{n+1}^- + \sigma_n^- \sigma_{n+1}^+\right)~,
\end{equation}
where the energies $\omega_n$ (in $H_{\rm TC}$) are random and uncorrelated, and with $J \ll W,g_c$. We find that even a very small hopping ($J/W=0.1$ with $g_c/W=10$) is enough to change all the properties of the CB, while keeping those of the GS and the FEB unaffected.

In Fig.~\ref{fig:Ps} it is shown that, when the integrability-breaking perturbation is added to the TC Hamiltonian, the level statistics in the CB changes from Poisson to Wigner-Dyson ({\bf a}), with level spacing ratio $\braket{r}=0.5(4)$, while in the FEB it remains semi-Poisson ({\bf b}). Moreover, for the CB the survival probability vanishes for $N \rightarrow \infty$ (see inset of Fig.~\ref{fig:EApar}{\bf a}), and the eigenstates become ergodic (see Fig.~\ref{fig:EApar}{\bf d}). As seen in Fig.~\ref{fig:EApar}{\bf d}, the multifractality of the GS and the FEB remain unaffected by nearest-neighbor hopping. The same robustness to nearest-neighbor hopping has been observed in the noninteracting case~\cite{schachenmayerPRB2020}, and it is consistent with the fact that the FEB reflects ``single-particle physics'' as its eigenstates are constituted by only one dark state.

\section{Possible experimental realization with cold atoms}

We propose to explore the multifractality of the interacting disordered Tavis-Cummings model using ultracold atoms trapped in a high-finesse cavity. To be concrete, consider $^{87}$Rb as in Ref.~\cite{longNP2023} with the spins encoded using the $|F=1,m_F=1\rangle$ and $|F=2,m_F=2\rangle$ levels. Circularly polarized light at $795$ nm incident from the side, either forming an incommensurate lattice~\cite{brantutNP2023} or a speckle pattern, will induce a random effective magnetic field due to the vector light-shifts with tunable strength. Raman coupling using a cavity photon then provides a realization of the disordered TC model with a tunable $g_c$~\cite{thompsonN2012}.

The dynamics will then be studied by initializing the system in a coherent superposition of excitations at the equator of the collective Bloch sphere (close to half filling) and letting the system evolve. The last generation of cavity experiments allows for local monitoring of the state of individual spins~\cite{stamper-kurnPRL2022,zhangPRL2023} and deterministic atom number preparation up to mesoscopic samples containing tens of particles, such that the spatial structure of the spin state can be directly observed. In particular, the participation ratios $I_q$ are experimentally accessible as a function of time and atom number. Even without local monitoring capabilities, it was shown in Ref.~\cite{brantutNP2023} that the participation ratio in the linear regime can be estimated from response functions, and it would be interesting to explore possible extensions to the interacting case.

The main limitation of such implementations is set by the finite photon lifetime, which yields an upper bound to the accessible timescales during which the system has a fixed number of excitations. Very narrow cavities with Rb atoms in the strong coupling regime have been demonstrated with photon lifetimes up to $35\,\mu$s~\cite{hemmerichS2012}. Beyond this timescale, continuous pumping would be needed to stabilize the system around an average number of excitations. The use of long-lived atomic states will ensure that dissipation is predominantly due to photons leaking from the cavity, which can then be detected.

Last, this experimental implementation suggests the study of the interacting Tavis-Cummings model in the presence of dissipation. In particular, the possible modification of the superradiant emission expected for fully delocalized excitations~\cite{temnov2005superradiance} may bear the signature of multifractality in our system, which will be the focus of a future work.

\section{Discussion and outlook}

The results above imply that  eigenfunctions are partially delocalized even if the level statistics suggests otherwise. Moreover, cavity-mediated interactions between excitons favor delocalization of the wavefunctions both in the integrable and nonintegrable models (see Figs.~\ref{fig:IPRq} and \ref{fig:EApar}{\bf d}). In the former, this is associated with a change in the exponent $\tau_q$ which is positive for $q>1$, as opposed to the frozen phase observed in the noninteracting case (see Fig.~\ref{fig:IPRq}{\bf a}). As a consequence, the many-body interactions make the eigenfunctions more delocalized, i.e., more similar to those of a good conductor.  The effects of interactions are  also striking in the nonintegrable case as they appear to turn the system into a perfect conductor in the middle of the spectrum. This is in stark contrast to the single-excitation model, where multifractality is robust to nearest-neighbor hopping perturbations at all energies~\cite{schachenmayerPRB2020}.

Those described above are examples of qualitative modifications in the many-particle eigenfunctions due to collective light-matter coupling of spins to a cavity field, resulting in new transport properties and phases of matter---a result relevant to the growing field of light-modified quantum matter~\cite{ebbesenS2021,rubioNRC2018}. They might find direct applications in improving exciton conductivity in, e.g., disordered molecular systems, where it has recently been observed that strong collective light-matter coupling can boost the diffusion coefficient, even inducing ballistic transport~\cite{schwartzNM2023,ebbesenAN2020}. They are also directly relevant for the latest generation of cavity quantum electrodynamics experiments~\cite{zhangPRL2023,stamper-kurnPRL2022} where the dynamics of few-atoms systems can be explored beyond the linear regime.

\begin{acknowledgments}

We acknowledge useful discussions with Jakob Reichel, Romain Long, Dmitry Abanin, Alexandre Faribault, Ivan Khaymovich, and Felipe Herrera. We thank especially Alexander Mirlin for clarifying the analytical estimate of the multifractal exponent. Numerical code for this work has been written in Julia~\cite{shahSR2017}, using QuantumOptics.jl~\cite{ritschCPC2018}. This work has benefited from state grants managed by the French National Research Agency under the Investments of the Future Program with the Grants No.~ANR-21-ESRE-0032 (aQCess), and No.~ANR-22-CE47-0013 (CLIMAQS). We acknowledge support from ECOS-CONICYT through Grant No.~C20E01 and the Centre National de la Recherche Scientifique (France) through the IEA 2020 campaign. Computing time was provided by the High Performance Computing Center of the University of Strasbourg. J.P.B.~acknowledges funding from the Swiss National Science Foundation (Grant No.~184654) and the Swiss State Secretariat for Education, Research and Innovation (Grant No.~MB22.00063).
\end{acknowledgments}

\footnotetext[1]{The level spacing ratio of the semi-Poisson distribution [$\braket{r}=0.613(7)$] has been computed by generating $10^6$ uniformly distributed random numbers and collecting the middle points of each pair of neighboring numbers. The Poisson value ($2\ln 2 -1$) has been extracted from the literature~\cite{husePRB2007,laflorencieCRP2018,santosPRL2020}.}

\bibliographystyle{apsrev4-2}
\bibliography{biblio}

\end{document}